# High Efficiency and ultra broadband optical parametric four-wave mixing in chalcogenide-PMMA hybrid microwires

Raja Ahmad[1,*] and Martin Rochette[1]

[1] *Department of Electrical and Computer Engineering, McGill University, Montreal (QC), Canada, H3A 2A7.*
[*]*raja.ahmad@mail.mcgill.ca*

**Abstract:** We present polymer (PMMA) cladded chalcogenide ($As_2Se_3$) hybrid microwires that realize optical parametric four-wave mixing (FWM) with wavelength conversion bandwidth as broad as 190 nm and efficiency as high as 21 dB at peak input power levels as low as 70 mW. This represents 3-30 x increase in bandwidth and 30-50 dB improvement in conversion efficiency over previous demonstrations in tapered and microstructured chalcogenide fibers with the results agreeing well with the simulations. These properties, combined with small foot-print (10 cm length), low loss (<4 dB), ease of fabrication, and the transparency of $As_2Se_3$ from near-to-mid-infrared regions make this device a promising building block for lasers, optical instrumentation and optical communication devices.

## 1. Introduction

The recent development of devices based on novel nonlinear materials like chalcogenides (ChGs), silicon (Si) and other semi-conductors has revolutionized the field of nonlinear photonics [1-3]. Among the nonlinear effects observed in these materials, four-wave mixing (FWM) is the process that finds the most applications including wavelength conversion [4], optical regeneration [5, 6], optical delay [7], time-domain demultiplexing [8], temporal cloaking [9] and negative refraction [10]. Of particular interest is the degenerate form of FWM (DFWM) in which two photons provided by a pump wave convert into a Stokes and an anti-Stokes photon. This leads to the simultaneous process of converting a signal from Stokes (or anti-Stokes) wavelength to the anti-Stokes (or Stokes) wavelength, and the amplification of the input signal. In the best DFWM conditions, the newly generated signal –called the idler– not only represents a wavelength-converted version of the input signal but it can even be more powerful than the original input signal. Although FWM has been observed in several media including chalcogenides [11-14], silicon [15, 16], bismuth [17] and silica [18-20], there is a continued quest for devices that realize efficient and broadband FWM while offering compactness, low-power consumption and compatibility with optical fibers at low insertion loss. Among the commonly used nonlinear materials, arsenic triselenide ($As_2Se_3$) chalcogenide glass boasts the highest nonlinear refractive index coefficient $n_2=2.3\times10^{-13}$cm$^2$/W [21], that is up to 1000× that of silica, 20× that of $Bi_2O_3$, 4× that of $As_2S_3$, and 3× that of Si [21-23].

Despite the large value of $n_2$ in $As_2Se_3$, the material exhibits high normal chromatic dispersion in the 1,550 nm wavelength band which, as explained below, prevents FWM in bulk medium or in large core optical fibers with low refractive index contrast cladding. This can however be remedied by stretching the $As_2Se_3$ fibers into microwires for which the anomalous waveguide dispersion overcomes the normal material dispersion. Such microwires also exhibit large values of waveguide nonlinear coefficient $\gamma$ (= $n_2\omega_P/cA_{eff}$, $\omega_P$ being the pump angular frequency, $c$ being the speed of light and $A_{eff}$ being the effective mode area in the microwire–) which lowers the required power threshold for nonlinear processes. The highest reported value of $\gamma$ in such microwires is more than 5 orders of magnitude larger than in silica fibers [24]. DFWM has been observed in $As_2S_3$ (air-clad) microwires but the process had improper phase-matching and led to low conversion efficiency (~20 dB) and narrow FWM gain bandwidth [11]. Incidentally, FWM has never been observed in microwires made of the most nonlinear chalcogenide glass, $As_2Se_3$.

In this work, we utilize $As_2Se_3$ microwires coated with poly methyl meth acrylate (PMMA) to attain phase-matching and generate the most efficient and broadband DFWM ever reported in such compact and power efficient devices. The PMMA cladding, in addition to optimizing the optical performance, imparts microwires remarkable physical strength [25], which otherwise are extremely fragile. Experiments were performed with wires of various core wire diameters to observe DFWM conversion efficiency as high as 21 dB and a 12 dB wavelength conversion range in extend of 190 nm, limited by the tunability range of available probe lasers. The large nonlinearity, the reduced chromatic dispersion from PMMA cladding and the long effective length (due to low absorption loss α < 1 dB/m – see Appendix at the end for further details on FWM performance comparison of various materials including silica, bismuth, silicon and chalcogenides –) of the hybrid microwires make them the most power efficient FWM devices, providing net broadband gain at peak pump powers as low as 70 mW.

## 2. Theory

In theory, the efficiency of DFWM process depends on phase-matching conditions given by [26]

$$\Delta k = 2\gamma P_P - \Delta k_L \quad (1)$$

where $P_P$ is the peak pump power, $\Delta k_L$ is the linear phase-mismatch that is chromatic dispersion dependent and is approximated by

$$\Delta k_L = -\beta_2(\Delta\omega)^2 - \tfrac{1}{12}\beta_4(\Delta\omega)^4 \quad (2)$$

where $\beta_i$ is the $i$-th order dispersion coefficient and $\Delta\omega$ is the angular frequency mismatch between the Stokes and anti-Stokes signals. It is well known that only the even order dispersion coefficients contribute to the phase-mismatch because of the symmetry in FWM processes [26]. The DFWM gain coefficient $g$ is then given by [26, 27]

$$g = \sqrt{(\gamma P_P)^2 - (\Delta k/2)^2} = \sqrt{\gamma P_P \Delta k_L - (\Delta k_L/2)^2} \quad (3)$$

In the case where the pump depletion is neglected as it transfers energy to the Stokes and anti-Stokes signals, the peak conversion efficiency $G_i$ of the idler wave is given by

$$G_i = P_{I,out}/P_{S,in} = (\gamma P_p/g)^2 \sinh^2(gL_{eff}) \quad (4)$$

where $P_{I,out}$ is the peak idler power at the output, $P_{S,in}$ is the input signal power and $L_{eff}$ is the interaction length. From Eq. (3), the gain is maximized when $\Delta k = 0$. This illustrates that in order to observe any DFWM gain, the linear dispersive phase-mismatch $\Delta k_L$ must lie within the range $0 < \Delta k_L < 4\gamma P_P$, where the upper limit is adjusted by the pump induced nonlinear phase shift. Although the influence of $\beta_4$ is expected to be much smaller than that of $\beta_2$, it becomes the dominant dispersion term when the pump lies close to zero-dispersion wavelength (ZDW), where $\beta_2 \sim 0$. Indeed, the DFWM is optimal both in terms of efficiency and bandwidth when the pump lies at the ZDW [26]. According to Eqs. (1-2) and ref. 26 however, both efficiency and bandwidth are reduced when $\beta_4$ is large and/or positive at the ZDW. Therefore, in order to get $g > 0$ over a broad bandwidth, in addition to $\beta_2 \sim 0$, the value of $\beta_4$ must be small and negative so that the $\Delta k$ approaches 0 with a small input pump power ($P_P > 0$). Figs. 1 (a) and (b) show the values of $\beta_2$ and $\beta_4$ in $As_2Se_3$ microwires surrounded by air or by a PMMA cladding, with a pump laser at $\lambda_P = 1,536$ nm. The $\beta_2$ and $\beta_4$ values are derived from the waveguide effective index neff which was numerically calculated from the characsersitic mode equation [28] using the refractive index parameters of $As_2Se_3$ provided in ref. [29] and those of PMMA in ref. [30].

Figs. 1 (a) and (b) show that in both situations whether the microwire is coated with air or PMMA, the $\beta_2$ profiles have two points where the chromatic dispersion is zero. Those are identified as ZDD 1 and ZDD 2 in order of decreasing wire diameter. Coating the $As_2Se_3$ microwire with PMMA (refractive index ~1.467 at $\lambda = 1,536$ nm) influences the values of $\beta_2$ and $\beta_4$ towards better phase-matching condition and more efficient-broadband DFWM gain with respect to the uncoated case. Comparing the $\beta_4$ value at larger zero-disperion diameters (ZDD $1_{\text{Air clad}} = 1.17$ µm and ZDD $1_{\text{PMMA clad}} = 1.013$ µm), it is found to increase from $\beta_{4,\text{ Air clad}}=-6.9 \times 10^{-6}$ ps$^4$/m to $\beta_{4,\text{ PMMA clad}}=-3.4 \times 10^{-6}$ ps$^4$/m by the application of a PMMA cladding rather than air. In addition, between the two ZDDs the value of $\beta_2$ –which is the dominant dispersion term in this range– advantageously reduces by up to one order of magnitude with the addition of the PMMA cladding. Finally, it is observed in Fig. 1 (c) that the value of $\gamma$ –calculated following the procedure ref. 31– increases from 76 W$^{-1}$m$^{-1}$ (at ZDD $1_{\text{Air clad}}$) to 98 W$^{-1}$m$^{-1}$ (at ZDD $1_{\text{PMMA clad}}$) from the addition of the PMMA cladding. This amounts to a total reduction in required pump power by 4.1 dB for compensating the linear phase mismatch when the pump lies at ZDD 1 and up to 10 dB when it lies between the two ZDDs – the range where the dispersion is anomalous. The PMMA cladding improves not only the DFWM efficiency of the device but it also lowers the power consumption.

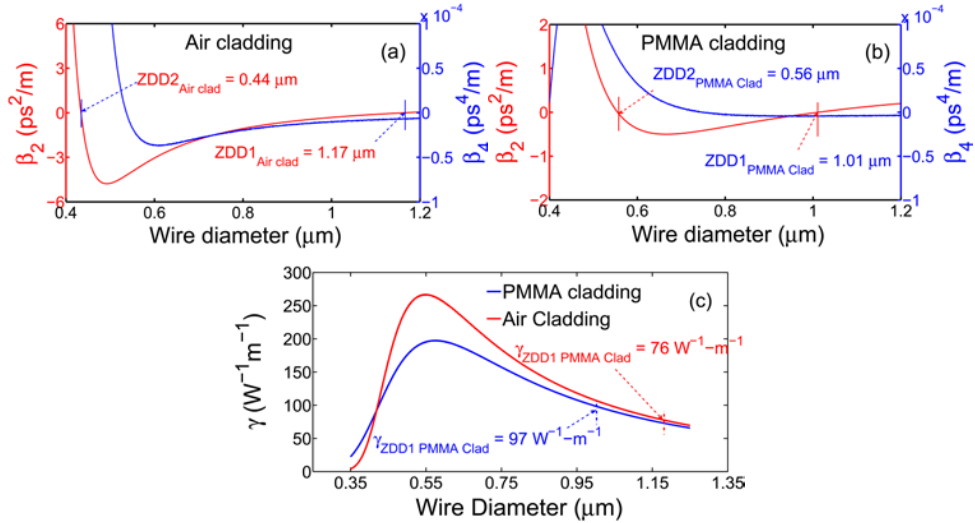

Fig. 1. Dispersion and Nonlinearity curves. Calculated values of second ($\beta_2$) and fourth ($\beta_4$) order dispersion profiles at $\lambda_P=1536$ nm, and the waveguide nonlinear coefficient ($\gamma$) as a function of $As_2Se_3$ wire diameter for the cases of air and PMMA cladding.

## 3. Experiment Design and Results

The hybrid As$_2$Se$_3$-PMMA microwires are fabricated using the procedure detailed in Ref. [24]. The microwires are pigtailed to standard silica fibers with a total insertion loss of <4 dB, a remarkably low value in comparison with similar waveguide structures [12,15,16]. The microwires support higher order fiber modes if the As$_2$Se$_3$ core diameter is greater than ~0.6 µm. However, by observing the modal profile at the output end using a camera and observing the mode interference patterns using an OSA, the input coupling was optimized towards the fundamental mode [32].

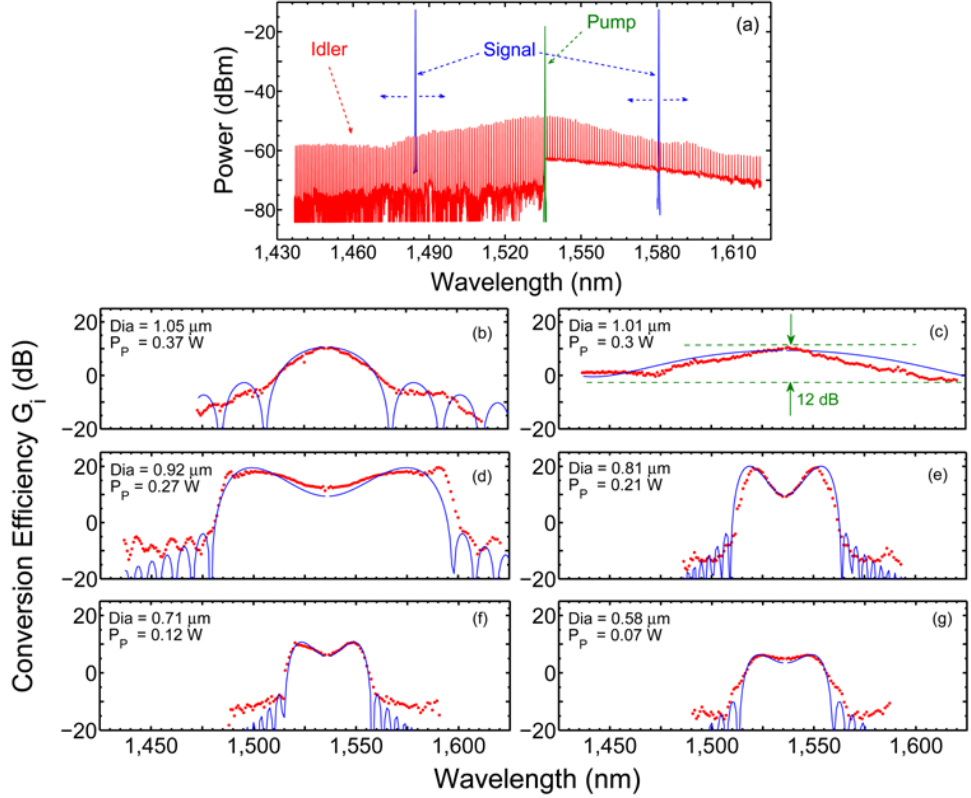

Fig. 2. DFWM output spectra and idler wavelength conversion efficiencies. (a) Optical spectra of the pump, signal and the generated idler corresponding to the signal tuned in the wavelength range 1460-1650 nm. The $As_2Se_3$ wire diameter was 1.01 µm, $P_P$ = 0.3 W, $P_{S,in}$ = -10.5 dBm and the microwire length was 10 cm. Only two sampled signal wavelengths are shown for clarity whereas one sample of idler spectrum per nm is given. (b)-(g) Idler conversion efficiencies spectra for a range of $As_2Se_3$ microwire diameters and a fixed length of 10 cm. The diameter value and the peak pump power used in the experiments are included as inset in each figure. Red (dotted) curves represent the experimental data, with the simulation fit denoted by the blue (continuous) lines.

The experimental measurement of DFWM in $As_2Se_3$-PMMA microwires is performed as follows: A pulsed pump laser centered at a wavelength of $\lambda_P$ = 1,536 nm and a co-polarized tunable continuous-wave (c.w.) laser signal were injected into the microwire. The pump laser generated transform limited pulses with a full width at half maximum (FWHM) duration of 5 ns and a repetition rate of 3.3 kHz (duty cylce ~ -48 dB). The resulting DFWM was observed at the microwire output using an optical spectrum analyzer (OSA). Fig. 2 (a) shows the spectra in response to a peak pump power $P_P$ = 0.3 W and an average signal power $P_{S,in}$ = -10.5 dBm at the input of a 1.01 µm diameter microwire. Two tunable laser sources with complementary wavelength coverage were used to characterize the idler wavelength conversion efficiency $G_i$ on either side of $\lambda_P$. The combined tuning range of the cw lasers was 1,460-1,650 nm and a corresponding idler was generated in the range 1,621-1,437 nm. It was observed that the pump remained undepleted in the process since the output pump power remained unchanged when the signal was turned on/off. In order to perform wavelength conversion efficiency measurements, the peak idler power $P_{I,\,out}$ at the output was derived from the average output idler power and the duty cycle of the pump laser. Using the $P_{I,\,out}$ and the input signal power $P_{S,\,in}$, the conversion efficiency was then calculated from the Eq. (4) provided above. The conversion efficiency measurements were performed with microwires of core diameters spreading in the range 0.58-1.05 µm that enabled covering the entire anomalous region between the two ZDDs. Figs. 2 (b)-(g) show the experimental results and corresponding simulations for those microwires. As shown in Fig. 2 (c), for a core diameter of 1.01 µm which approximately superimposes ZDD 1 and the pump wavelength, the value of $G_i$ reaches up to 10 dB and remains >-2 dB for a signal tuning range of 190 nm. The efficiency is expected to get even more flat with an increase in input pump power [26, 27]. The measured DFWM bandwidth is limited by the tuning range of the probe signal. Moreover, engineering of $\beta_2$ by the PMMA cladding allowed for a wide range of wire diameters, a broadband (≥ 50 nm) and efficient DFWM wavelength conversion (up to 21 dB efficiency) from a low peak power pump (70-370 mW). From a fabrication point of view, this adds tolerance by the DFWM process to a target microwire diameter. And from an application point of view, the bandwidth and gain of the DFWM device compare well with those of commercial fiber amplifiers doped with erbium and/or ytterbium [33]. In each of the conversion efficiency spectra, the impact of Raman scattering is also observed, appearing in the form of gain (loss) at a Stokes (anti-Stokes) wavelength shift of 7 THz or ~53 nm at 1550 nm with respect to the pump wavelength.

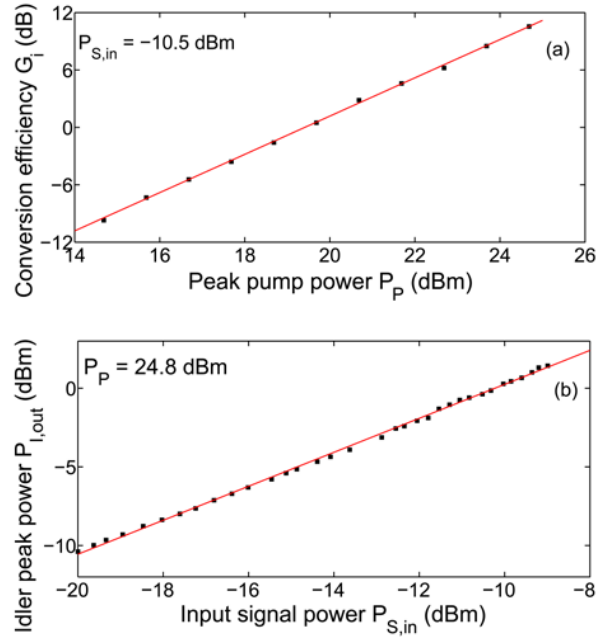

Fig. 3. DFWM device performance characteristics. (a) Measured idler conversion efficiency plotted as a function of the peak input pump power for a microwire with 1.01 µm diameter. (b) Measured idler peak power plotted as a function of input signal power for the same microwire.

Fig. 3 (a) depicts the idler conversion efficiency of the DFWM process in a microwire with 1.01 µm core diameter. The input signal power and the wavelength were maintained at -10.5 dBm and 1,540.8 nm, respectively. The plot shows a quadratic dependence of the conversion efficiency on peak pump power which corresponds to theory. There is no sign of saturation due to multiphoton absorption up to 0.3 W of input peak pump power which corresponds to the peak intensity of 37 MW-cm$^{-2}$. Fig 3(b) shows the dependence of the idler (peak) output power on the input signal power for a fixed (peak) pump power of 0.3 W. In this case, the idler output power increases linearly with the input signal power which again confirms the theoretical prediction. Since there is no sign of saturation, the conversion efficiency is expected to increase further from raising the input pump power. Likewise, the wavelength conversion efficiency can be enhanced further by employing longer microwires.

## 4. Summary

In summary, the addition of a PMMA cladding to an $As_2Se_3$ microwire advantageously enabled engineering the waveguide dispersion towards an optimization of DFWM process. This dispersion engineering accompanied by an enhanced nonlinear coefficient γ of the microwire (~ $10^5 \times$ that of silica fibers) led to an efficient and broadband DFWM in 10 cm long microwires at peak power levels as low as 70 mW. The next step is the introduction of resonant feedback [34] to realize a compact and low-threshold all-chalcogenide parametric oscillator. Such a compact and efficient device will find a wide-range of applications in telecommunication systems, mid-infrared spectroscopy and free-space communications [35].

**Acknowledgements**:
The authors gratefully acknowledge the support from Profs. Nicolas Godbout and Suzanne Lacroix at Ecole Polytechnique, Université de Montreal, Canada, for lending the pump laser used in experiments. We are also grateful to Dr. Phillippe de Sandro and Mr. Stephane Chatigny at Coractive High Tech inc, for providing chalcogenide fibers. This work was financially supported by FQRNT (Le Fonds Quebecois de la Recherche sur la Nature et les Technologies) and the Natural Sciences and Engineering Research Council of Canada (NSERC).

**Appendix**

This Appendix contains a detailed and quantitative comparison of the FWM performance of various optical materials including silica, bismuth, silicon, and chalcogenides ($As_2S_3$ and $As_2Se_3$). The appendix table. AT1 shows the values of nonlinear coefficient γ, propagation loss α and the calculated effective lengths for DFWM devices of the same foot-print (actual length L = 10 cm) but different materials including highly nonlinear (HNL) silica and bismuth fibers, $As_2S_3$ and Si waveguides and $As_2Se_3$ microwires. The large γ and low α values in $As_2Se_3$ microwires (and hence longer $L_{eff}$) leads to a much higher idler conversion efficiency $G_i$ (and parametric gain $G_S$) from the DFWM process, assuming a zero phase-mismatch ($\Delta k=0$). The values of $G_i$ and $G_S$ are plotted in Apendix Fig. AF1 for 10 cm long (fiber or planar) waveguides made from silica, bismuth, silicon and chalcogenides, under the assumption of perfect phase matching. This figure shows that As2Se3, because of lower absorption loss and high nonlinearity, is the best choice for realizing compact, power efficient FWM related devices.

**Supplementary table AT1**. Comparison of highly nonlinear fibers made from silica and bismuth, waveguide made from $As_2S_3$ chalcogenide and silicon and the microwires made from $As_2Se_3$ for a given foot-print of the devices.

|  | γ ($W^{-1}$-$m^{-1}$) | Loss | Actual | Effective length $L_{eff}$ |
|---|---|---|---|---|
| Highly nonlinear (HNL) silica fiber | 0.021 | $10^{-3}$ | 10 | 10.00 |
| Bismuth oxide ($Bi_2O_3$) fiber [AR1] | 1.36 | 0.8 | 10 | 9.91 |
| $As_2S_3$ waveguide [AR2] | 9.9 | 60 | 10 | 5.42 |
| Silicon waveguide [AR3] | 150 | 400 | 10 | 1.09 |
| $As_2Se_3$ microwire [This work] | 97 (at ZDD 1) | < 1 | 10 | 9.89 |
|  | 197 (at ZDD 2) | < 1 | 10 | 9.89 |

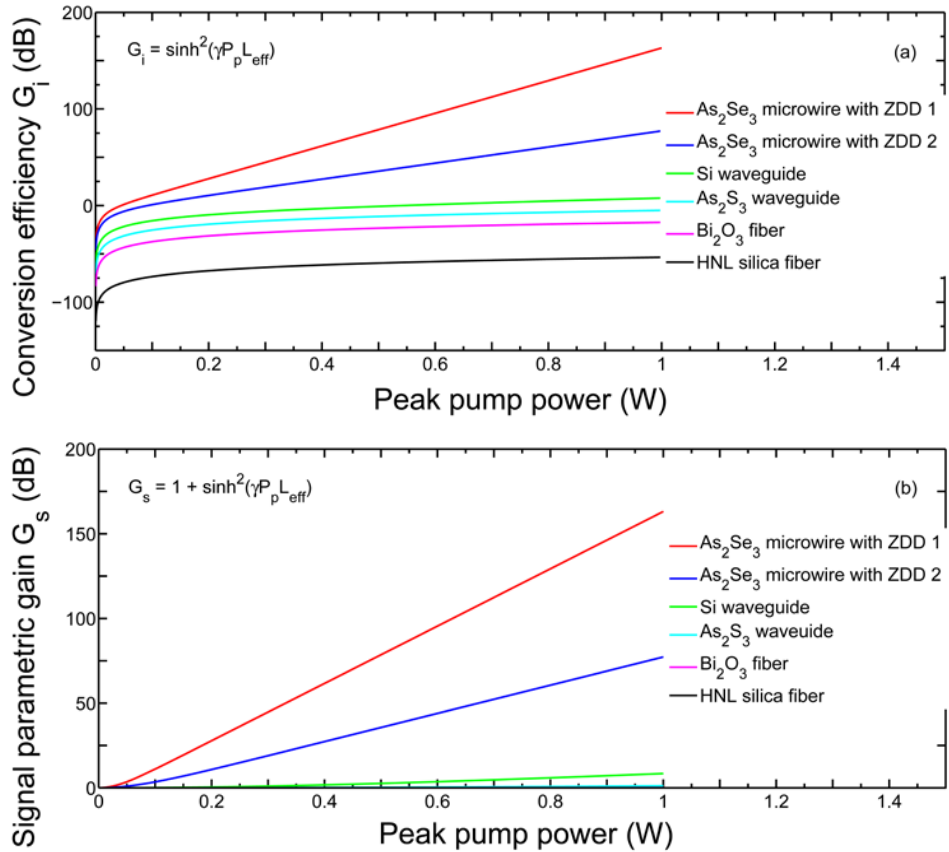

**Appendix fig. AF1.** Calculated (a) idler conversion efficiency and (b) signal gian for phase matched DFWM process in 10 cm long chalcogenides ($As_2Se_3$, $As_2S_3$), silicon, bismuth and silica fibers/waveguides. The low propagation loss with a high nonlinear coefficient in $As_2Se_3$ microwires allows the maximum conversion efficiency/signal gain in a compact scheme. The parameters used in these simulations are provided in the appendix table AT1 given above.

## Appendix References